\documentclass[journal=jacsat,manuscript=article]{achemso}

\usepackage{chemformula}
\usepackage[T1]{fontenc}
\usepackage{enumitem}
\usepackage{placeins }

\author{Moritz Engl}
\email{moritz.engl@namlab.com}
\affiliation[NaMLab gGmbH]{NaMLab gGmbH, 01187 Dresden}
\author{Wassim Hamouda}
\affiliation[HZB]{Helmholtz-Zentrum Berlin für Materialien und Energie, 14109 Berlin}
\author{Ines Häusler}
\affiliation[Institute of Physics]{Helmholtz-Zentrum Berlin für Materialien und Energie, 14109 Berlin}
\author{Suzanne Lancaster}
\affiliation[NaMLab gGmbH]{NaMLab gGmbH, 01187 Dresden}
\author{Luca Carpentieri}
\affiliation[NaMLab gGmbH]{NaMLab gGmbH, 01187 Dresden}
\author{Thomas Mikolajick}
\affiliation[NaMLab gGmbH]{NaMLab gGmbH, 01187 Dresden}
\alsoaffiliation[TU Dresden]{Chair of Nanoelectronics, TU Dresden, 01187 Dresden}
\author{Catherine Dubourdieu}
\affiliation[HZB]{Helmholtz-Zentrum Berlin für Materialien und Energie, 14109 Berlin}
\affiliation[Berlin]{Freie Universität Berlin, Physical and Theoretical Chemistry, 14195 Berlin}
\email{catherine.dubourdieu@helmholtz-berlin.de}
\author{Stefan Slesazeck}
\email{stefan.slesazeck@namlab.com}
\affiliation[NaMLab gGmbH]{NaMLab gGmbH, 01187 Dresden}

\title{Peak splitting and bias fields in ferroelectric hafnia mediated by interface charge effects}

\begin{document}

\begin{abstract}
The pristine state of hafnium based ferroelectric devices exhibits various unwanted properties, such as imprint and peak splitting, which diminish with bipolar cycling. The incorporation of a niobium oxide layer at different positions in metal-ferroelectric-metal and metal-ferroelectric-insulator-metal stacks is used to modify the pristine state of the device. X-ray photoelectron spectroscopy and transmission electron microscopy measurements are used to investigate the influence of niobium oxide on the zirconium hafnium oxide layer. It is hypothesized that the charged vacancies generated by the introduced niobium oxide in the adjacent zirconium hafnium oxide layer result in an electric bias field that influences the pristine polarization state of the domains. A comparison of different stacks shows that peak splitting in the pristine state is most likely related to the formation of opposing electric bias fields in upwards and downwards polarized domains.
Furthermore, the incorporation of niobium oxide in the zirconium hafnium oxide/aluminum oxide capacitor stack in between the ferroelectric and insulating layer leads to a peak splitting free device without imprint, which could be explained by the increased influence of charge trapping near the zirconium hafnium oxide-/niobium oxide and niobium oxide-/aluminum oxide interfaces.

\end{abstract}

\section{Introduction}
Ferroelectric domains in hafnium oxide layers have two stable operation points with opposite polarization which arises from two energetically stable configurations of the ferroelectric orthorhombic phase \cite{kim2019ferroelectric}. The dynamic IV-curve, measured using the PUND method (Positive Up Negative Down) \cite{scott1988}, exhibits the characteristic hysteresis with a specific remanent polarization. 
The initial state of ferroelectric capacitors based on hafnium oxide thin films usually exhibits multiple non ideal features in the IV-curve. These effects encompass for example an initial shift along the voltage axis of the hysteresis loop (imprint) \cite{jeong2022oxygen,lee2023analysis}, peak splitting and therefore pinching \cite{lee2023analysis,pevsic2016physical} of the hysteresis loop or wakeup (an increase of the remanent polarization with cycling \cite{pevsic2016physical}).
The cycling instability of the hysteresis impedes the application of these films in actual devices and circuits, because either an initial cycling of the device before the actual operation is necessary or the changes during operation have to be considered in the design process of the circuit.

Recently it has been shown that a higher amount of orthorhombic phase and less peak splitting can be observed in metal-ferroelectric-metal (MFM) capacitors by using a TiO\textsubscript{x} capping layer and a Nb\textsubscript{2}O\textsubscript{5} capping layer between the zirconium doped hafnium oxide (HZO) layer and the TiN electrodes \cite{walke2024doped,popovici2022high}. It is assumed that TiO\textsubscript{x} leads to a more favorable orientation of the grains in the HZO\cite{popovici2022high} and that the introduced Nb\textsubscript{2}O\textsubscript{5} layer acts as an oxygen source which reduces the amount of oxygen vacancies generated in the HZO\cite{popovici2022high}. 

In Popovici et. al. \cite{popovici2022high} the assumption that niobium oxide acts as an oxygen source is supported by ab initio calculations and the lower electron affinity of Nb\textsubscript{2}O\textsubscript{5} in comparison to HfO\textsubscript{2} and ZrO\textsubscript{2}, which indicates that the migration of oxygen ions from niobium oxide to the HZO is energetically more favorable.
Nevertheless, pure niobium can be used in Pt-Nb-HfO\textsubscript{x}-Pt structures as an oxygen exchange layer \cite{nandi2014resistive}, which facilitates the formation of vacancies in the hafnium oxide layer. The use of Nb as an oxygen exchange layer indicates that the formation of niobium oxide phases may be energetically favourable and therefore non-stoichiometric niobium oxide may lead to the formation of oxygen vacancies in the HZO layer.

In this paper the initial state of metal-ferroelectric-metal (MFM) and metal-ferroelectric-insulator-metal (MFIM) is investigated with the introduction of a thin niobium oxide layer at different positions in the stacks. The investigations are based on two basic kinds of stacks, a MFM stack with HZO as the ferroelectric layer and a MFIM stack with HZO as the ferroelectric and aluminum oxide (AlO\textsubscript{x}) as the insulating layer. The MFIM device with 4 nm AlO\textsubscript{x} is used as a standard layer to observe negative capacitance (NC) as described in Engl. et. al.\cite{engl2023degradation}.\\
Furthermore, X-ray photoelectron spectroscopy (XPS) and transmission electron microscopy (TEM) measurements are performed to investigate the chemical interaction between HZO and niobium oxide.

\section{Device structures}
\label{section_device_structure}
All devices were fabricated on p-doped silicon substrates. The bottom electrode was formed by sputtering 30\,nm of W and 10\,nm of TiN in high vacuum. Subsequently an ALD process at 280 degree with ozone as oxidant was used to deposit 10\,nm of zirconium hafnium oxide (HZO) as the ferroelectric layer. For the MFIM samples 4\,nm aluminium oxide (AlO\textsubscript{x}) was deposited by ALD with water as oxidant. The top electrode consists of 10\,nm TiN sputtered in high vacuum. An annealing step at 500 degrees Celsius for 20\,s in N\textsubscript{2} atmosphere was used to crystallize the HZO after the deposition of the top TiN layer. 

Subsequently 10\,nm of Ti and 25\,nm of Pt were evaporated and structured into rectangular dots in a lift-off process. The top TiN layer was etched by reactive ion etching with the use of the already structured Pt dots as etching masks. For the bottom electrode contact an additional lithography step was used followed by reactive ion etching of the ferroelectric/dielectric layer stack stopping on the tungsten electrode. 
The 1\,nm thick niobium oxide (NbO\textsubscript{x}) layer inserted in the different stacks was sputter deposited in high vacuum. The thickness of oxygen scavenging layers has been shown to be critical in either enhancing or degrading the ferroelectric properties \cite{lancaster2023toward}, and layers around 1\,nm are expected to be optimal for enhancing ferroelectricity, which may be dependent on the oxygen scavenging material.

The electrical properties of the device are influenced by the adjacent layers of the HZO \cite{alcala2023electrode,lee2021influence}. The induced tensile stress during the annealing \cite{lee2021influence} based on the difference in thermal expansion coefficients of the layers as well as the formation of interfaces \cite{alcala2023electrode} like titanium nitride oxide at the HZO-TiN interfaces impact the resulting distribution of phases in the HZO after the thermal anneal. It is assumed that an oxidation process\cite{alcala2023electrode,hsain2022role} during the anneal is responsible for the interfacial layer(s) generation by scavenging partially the necessary oxygen from the HZO layer and therefore generating oxygen vacancies within the HZO layer in proximity of the interface(s)\cite{alcala2023electrode,lee2021influence,hsain2022role,hamouda2020physical,hamouda2022oxygen}.
Considering the process flow of a MFM device with TiN as top and bottom electrode, the formation of an initial titanium nitride oxide layer is expected during the ALD process and the transfer process between the deposition tools which impedes the scavenging of oxygen during the annealing step \cite{hamouda2020physical,hsain2022role}, resulting in the formation of fewer vacancies at the bottom interface than at the top interface. In the case of a NbO\textsubscript{x} layer below the HZO layer, the NbO\textsubscript{x} layer is assumed to be more oxidized during the ALD deposition than the NbO\textsubscript{x} in contact with the layer at top electrodes, similar to the investigations with tantalum oxide in Shin et al. \cite{shin2023method}.\\
The introduction of oxygen vacancies is associated with a higher orthorhombic phase fraction and thus a higher remanent polarization Pr, but is also considered to reduce device endurance.\cite{alcala2023electrode}. Furthermore, the redistribution and the additional formation of vacancies during cycling can be associated to the changes of device properties\cite{lee2023analysis,pevsic2016physical}. 

\begin{figure}[!htb]
\includegraphics[width=6.0in]{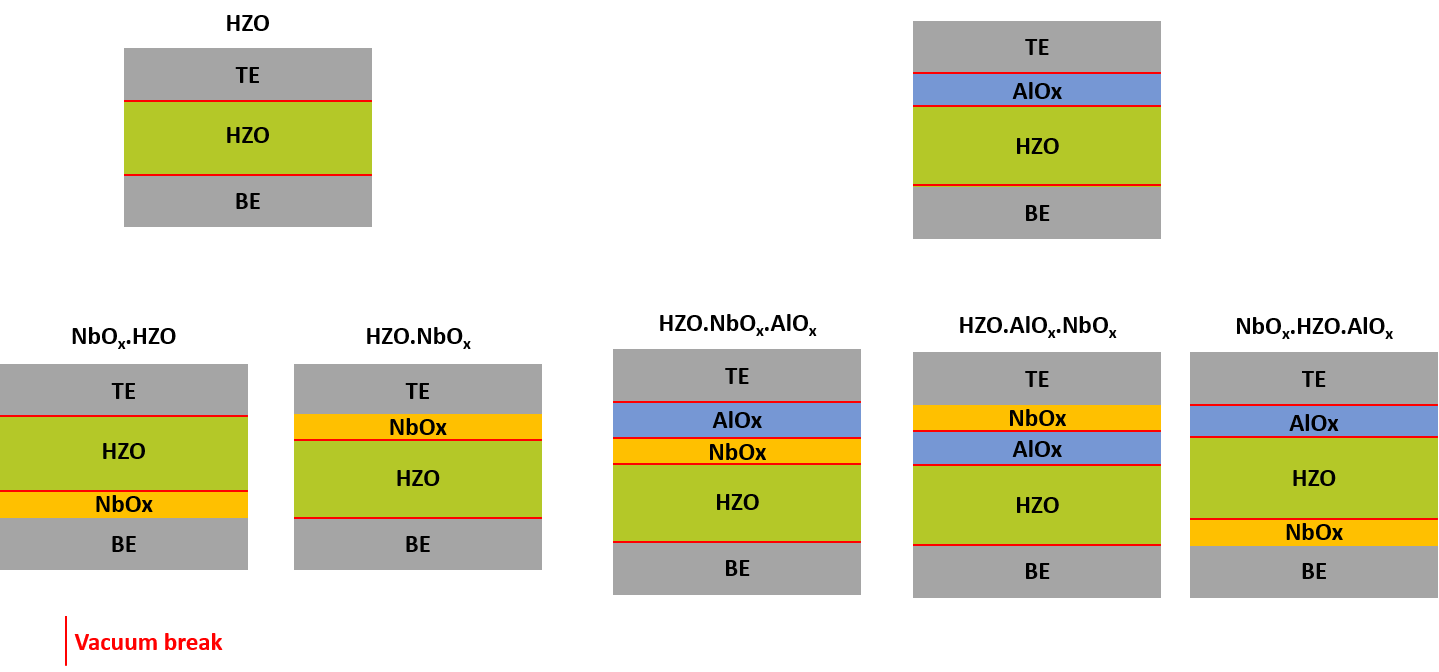}
\caption{Schematic cross section of the investigated samples. TE is the top electrode and BE the bottom electrode. The name above each stack is used as a short name for the respective sample. A red line indicates that a vacuum brake during processing occurred.}
\label{fig:samples}
\end{figure}

\begin{figure}[!htb]
\includegraphics[width=5.0in]{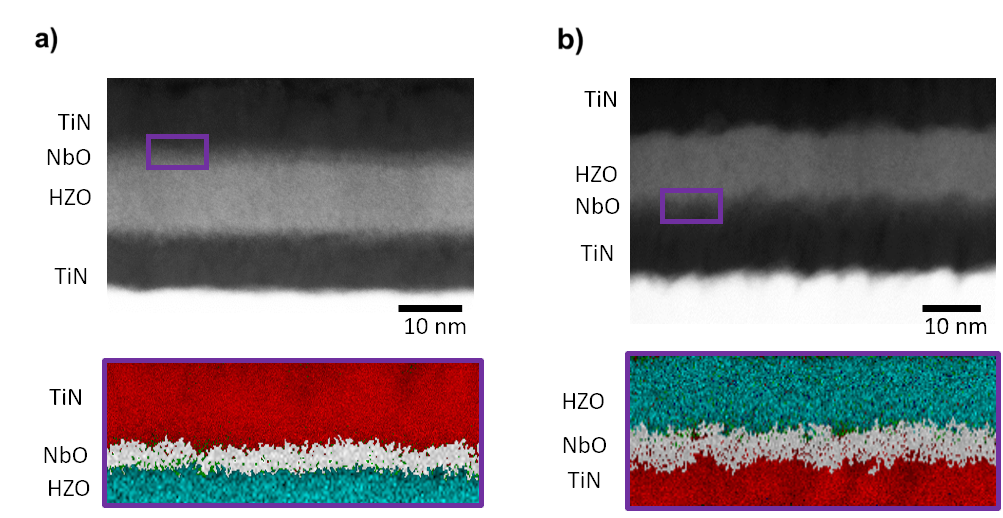}
\caption{HAADF-STEM and EELS measurements of two samples HZO.NbO\textsubscript{x} in a) and NbO\textsubscript{x}.HZO in b). Each color in the magnified images in the bottom represents a signal from a specific element: red-N, grey-Nb, cyan-Hf and green-O. See Fig. S1 in the supplementals for more information.}
\label{fig:TEM}
\end{figure}

\begin{figure}[!htb]
\includegraphics[width=5.0in]{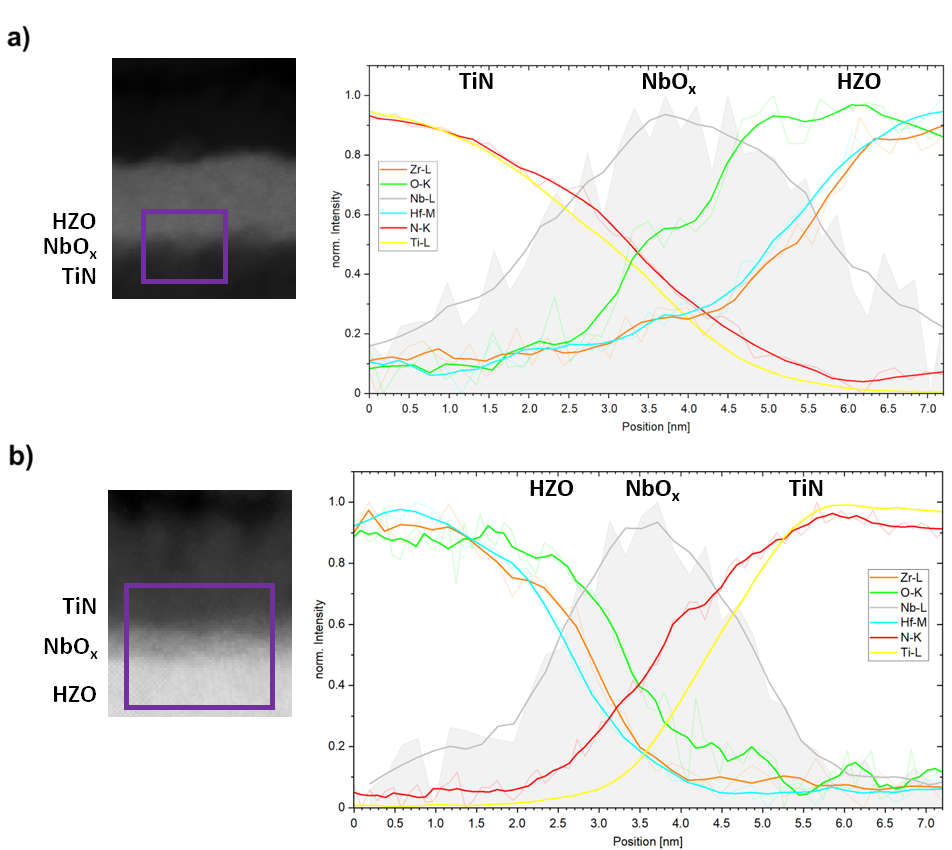}
\caption{Elemental analysis for the TiN-NbO\textsubscript{x}-HZO region of a NbO\textsubscript{x}.HZO device in a) and the HZO-NbO\textsubscript{x}-TiN region of a HZO.NbO\textsubscript{x} device in b). The purple squares mark the region, where the analyses were carried out.}
\label{fig:diffusion}
\end{figure}

Overall seven device types were manufactured for electrical measurements and analysis, as depicted in Fig. \ref{fig:samples}. 
The HAADF-STEM and EELS measurements in Fig. \ref{fig:TEM} show a comparable high roughness of the layers, as expected for polycrystalline layers, which has to be taken into account in the further analysis, since each measured signal in the EELS spectra is the superposition of several contributing atomic layers. 
The NbO\textsubscript{x}.HZO sample (Fig. \ref{fig:diffusion}(a)) shows a broader distribution of the niobium signal and a higher oxygen signal near the NbO\textsubscript{x}-/HZO interface compared to the ones in the HZO.NbO\textsubscript{x} sample (Fig. \ref{fig:diffusion}(b)). The larger oxygen content is attributed to the exposure of the NbO\textsubscript{x} layer to ozone in the first steps of the HZO growth. The broader Nb distribution is either caused by the diffusion of niobium in the HZO and/or TiN layer or it is a result of a difference in the roughness between the NbO\textsubscript{x} layers in Fig. \ref{fig:diffusion} a) and b). The HZO.NbO\textsubscript{x} in Fig. \ref{fig:diffusion} b) shows a higher nitrogen content in the NbO\textsubscript{x} layer than the NbO\textsubscript{x}.HZO sample in Fig. \ref{fig:diffusion} a), which is attributed to the exposure of the NbO\textsubscript{x} layer in the HZO.NbO\textsubscript{x} sample to the nitrogen gas used in the reactive sputtering of TiN with a titanium target. 

\FloatBarrier
The XPS investigations required the manufacturing of different samples, since a measurement through the 20\,nm thick Pt layer that defines the TE of the capacitor structure is not possible and the surface next to dots is most likely damaged by the dry etching of the TiN layer. For this experiment an alternative process flow was used. The deposition of TiN at the top electrode was skipped and the structuring of the Ti/Pt electrodes was done by a shadow mask. The contact to the bottom electrode for electric measurements was realized by electrically-broken capacitors which form an ohmic contact to the bottom electrode.
The electrical properties of these samples are expected to be different from those annealed with a top TiN layer. Two samples, one HZO.AlO\textsubscript{x} sample and one HZO.NbO\textsubscript{x}.AlO\textsubscript{x} sample, were manufactured. The HZO and AlO\textsubscript{x} layers for the XPS experiments were deposited in the Oxford Instruments OpAL system.\\
The HZO and AlO\textsubscript{x} layers for the electrical analysis and TEM measurements were deposited in the Scia Atol 200 tool from Scia Systems. The Opal system used HyALD and ZyALD as precursors, while the Scia system used TEMAHf as precursor. Despite having a similar HZO stoichiometry of Hf\textsubscript{0.4}Zr\textsubscript{0.6}O\textsubscript{2}, the Scia samples exhibits more peak splitting in the pristine state, making it easier to study phenomena such as peak splitting in this particular experiment.

\section{XPS investigations}
XPS measurements were performed on HZO.AlO\textsubscript{x} and HZO.NbO\textsubscript{x}.AlO\textsubscript{x} samples without top TiN layers. 
The concentration of oxygen vacancies, which are assumed to be positively charged, near the top of the HZO layer was calculated from the intensities $I_{Hf3+}$ of the Hf\textsuperscript{+3} state and $I_{Hf4+}$ of the Hf\textsuperscript{4+} state of the 4f orbital of the hafnium atoms, assuming that the formation of oxygen vacancies changes the charge state of the Hf atoms. 
\begin{equation}
    V_{\Ddot{O}} = \frac{I_{Hf3+}}{I_{Hf3+}+I_{Hf4+}} \cdot0.5\cdot0.25
    \label{eq_Vo_conc}
\end{equation}

\begin{figure}[H]
\includegraphics[scale=0.6]{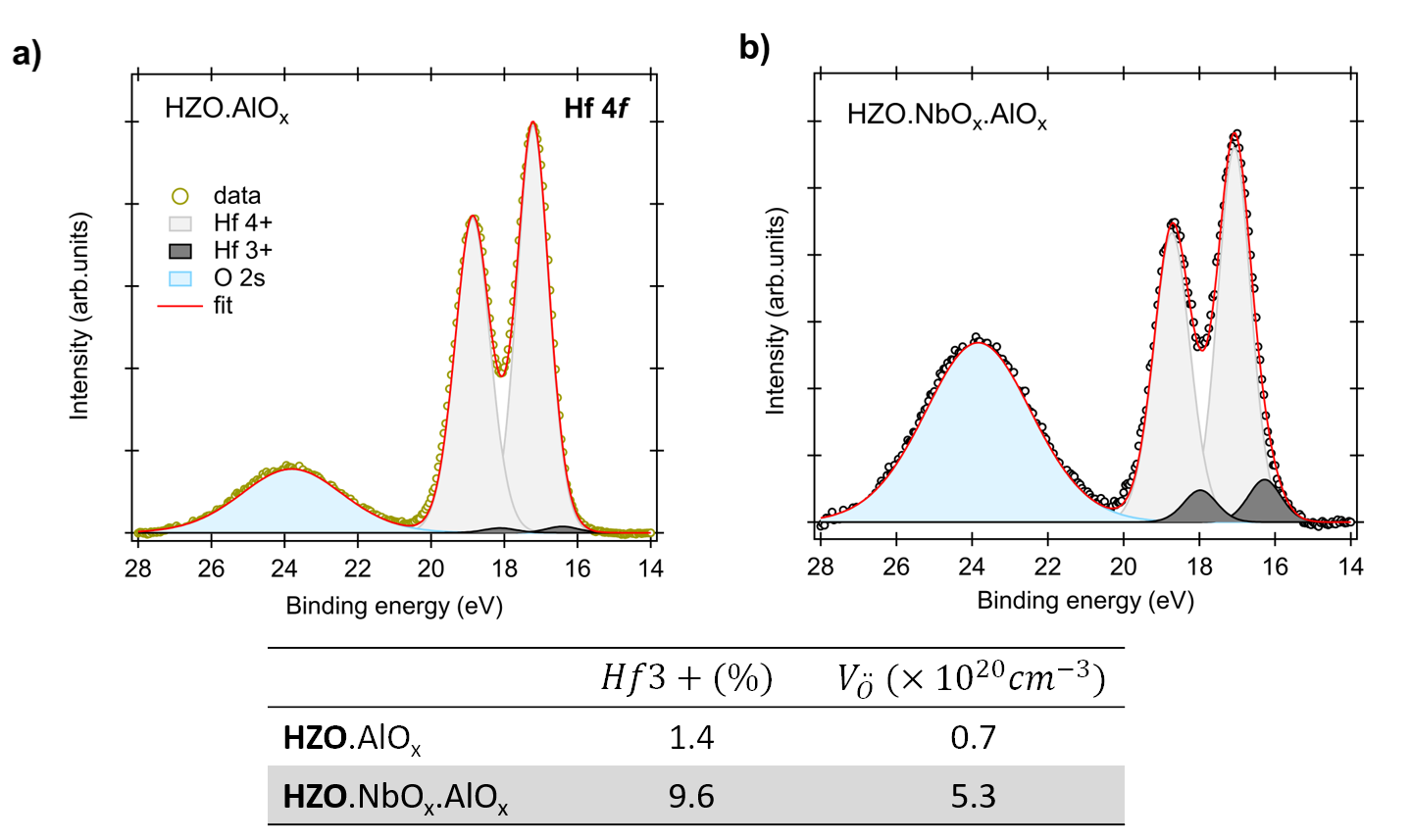}
\caption{XPS measurements on the Hf 4f and O 2s core levels for the HZO.AlOx device in a) and the HZO.NbO\textsubscript{x}.AlO\textsubscript{x} device in b). The red curve is the fit of the experimental data. In the table below the graphes, we give the intensity of the Hf3+ peak and the concentration of the oxygen vacancies calculated using equation \ref{eq_Vo_conc}.}
\label{fig:concentration_vacancy}
\end{figure}

The density of vacancies in the HZO.NbO\textsubscript{x} sample is of 5.3$\times$10\textsuperscript{20}\,cm\textsuperscript{-3}, which is higher than the density of vacancies in the HZO.AlO\textsubscript{x} sample of 0.7$\times$10\textsuperscript{20}\,cm\textsuperscript{-3}, as shown in Fig. \ref{fig:concentration_vacancy}.

The increase of oxygen vacancies in HZO by the insertion of the NbO\textsubscript{x} layer on top indicates that NbO\textsubscript{x} acts as an oxygen scavenger, thus increasing the amount of vacancies in the HZO layer. 
It should be noted that the NbO\textsubscript{x} in this work is sputtered while the NbO\textsubscript{x} in the literature \cite{popovici2022high,walke2024doped} is deposited by ALD, hence a difference in composition and defects is expected. The different behavior of the NbO\textsubscript{x} is most likely related to a different stoichiometry after depositing the layer. The tendency to scavenge oxygen indicates that the NbO\textsubscript{x} layer is not fully saturated with oxygen after deposition on top of HZO, and that oxygen is scavenged from neighboring layers to get closer to the most energetically favorable stoichiometry.\\
According to theoretical calculations by Jacob et. al. \cite{jacob2010thermodynamic}, the introduction of defects in Nb\textsubscript{2}O\textsubscript{5-x} leads to a more stable composition than pure Nb\textsubscript{2}O\textsubscript{5}. 
An analysis of the NbO\textsubscript{x} composition based on the 3d core level of Nb is beyond sensitivity, which is concluded from the deviation between the measurement signal and the fit in Fig. S2 of the supplementary. The deviation could indicate a strong inter-diffusion of the materials, which can't be estimated due to the roughness of the layers, a high amount of defects in the NbO\textsubscript{x} or the formation of substoichiometric oxide phases.   

Furthermore, the analysis of the AlO\textsubscript{x} layers in the HZO.AlO\textsubscript{x} and HZO.NbO\textsubscript{x}.AlO\textsubscript{x} samples show, that the amount of stoichiometric Al\textsubscript{2}O\textsubscript{3} decreases from 97\,\% in the HZO.AlO\textsubscript{x} sample to 89\,\% in the HZO.NbO\textsubscript{x}.AlO\textsubscript{x} one. Therefore it can be assumed that NbO\textsubscript{x} scavenges oxygen from AlO\textsubscript{x} when in direct contact with it. A detailed analysis is shown in the supplementary Fig. S3.\\
An alternative explanation for the change of stoichiometric Al\textsubscript{2}O\textsubscript{3} between the two samples is that the AlO\textsubscript{x} layer scavenges oxygen from the HZO layer when in direct contact with it. The introduction of the NbO\textsubscript{x} layer between the HZO and AlO\textsubscript{x} acts as a barrier to this scavenging process. The presence of 7.6 times more charged vacancies in the HZO.NbO\textsubscript{x}.AlO\textsubscript{x} sample than in the HZO.AlO\textsubscript{x} one (Fig. \ref{fig:concentration_vacancy}) indicates that NbO\textsubscript{x} exhibits a greater tendency to scavenge oxygen compared to AlO\textsubscript{x}. Therefore, the NbO\textsubscript{x} layer most likely scavenges oxygen from the neighboring AlO\textsubscript{x} layer and the neighboring HZO layer.

The electrical results for the two XPS samples are shown in the supplementary Fig. S4. They show similar trends to the other samples comprising the insertion of a NbO\textsubscript{x} layer, which are discussed in the next section.

\section{Electrical analysis of MFM with NbO\textsubscript{x} incorporation}
The electrical characterization was performed using a Keithley 4200A-SCS Parameter Analyzer in combination with a 4225 Pulse Measurement Unit (PMU) and 4225 Remote Preamplifier/Switch Modules (RPMs). For each measurement the bottom electrode was grounded and the pulses were applied to the top electrode of the device. The current was measured at the grounded bottom electrode to reduce the influences of parasitic capacitances from the measurement setup.
The ferroelectric properties were analyzed by PUND\cite{scott1988} measurements after bipolar cycling as depicted in Fig. S5 of the supplementary. The cycling and measurement amplitudes for MFM devices were 3\,V and for MFIM devices 7.5\,V to take the voltage drop over the AlO\textsubscript{x} and or NbO\textsubscript{x} layers into account.

\begin{figure}[H]
\includegraphics{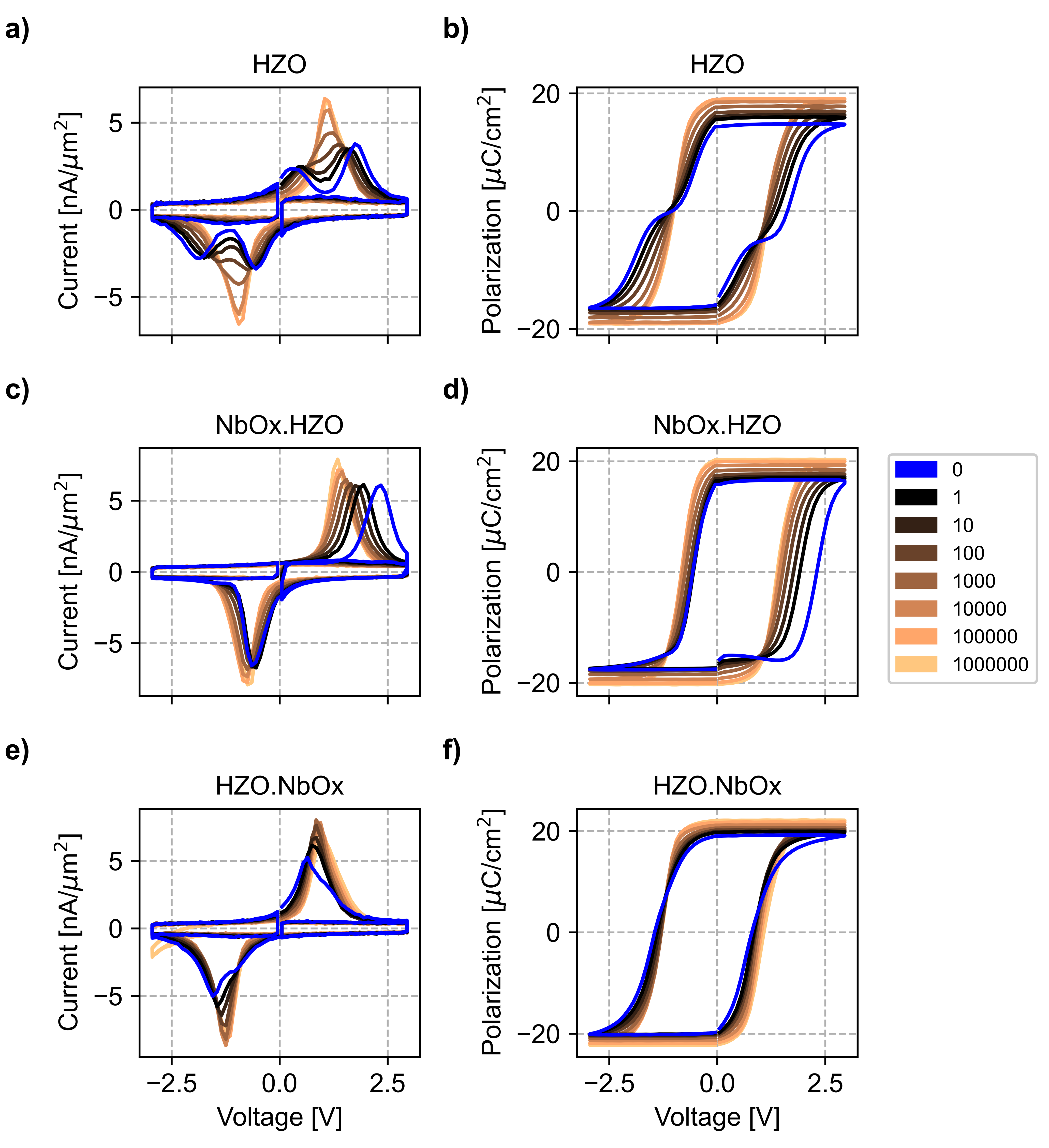}
\caption{IV- and PV-curves from PUND measurement with increasing cycling number for the investigated HZO sample in a) and b), the NbO\textsubscript{x}.HZO sample in c) and d) as well as the HZO.NbO\textsubscript{x} sample in e) and f). For the IV-curves the current densities vs the applied voltages are plotted and for the PV loops the difference of P-U and N-D pulses are used to estimate the polarization.}
\label{fig:IV_MFM}
\end{figure}

\begin{figure}[H]
\includegraphics{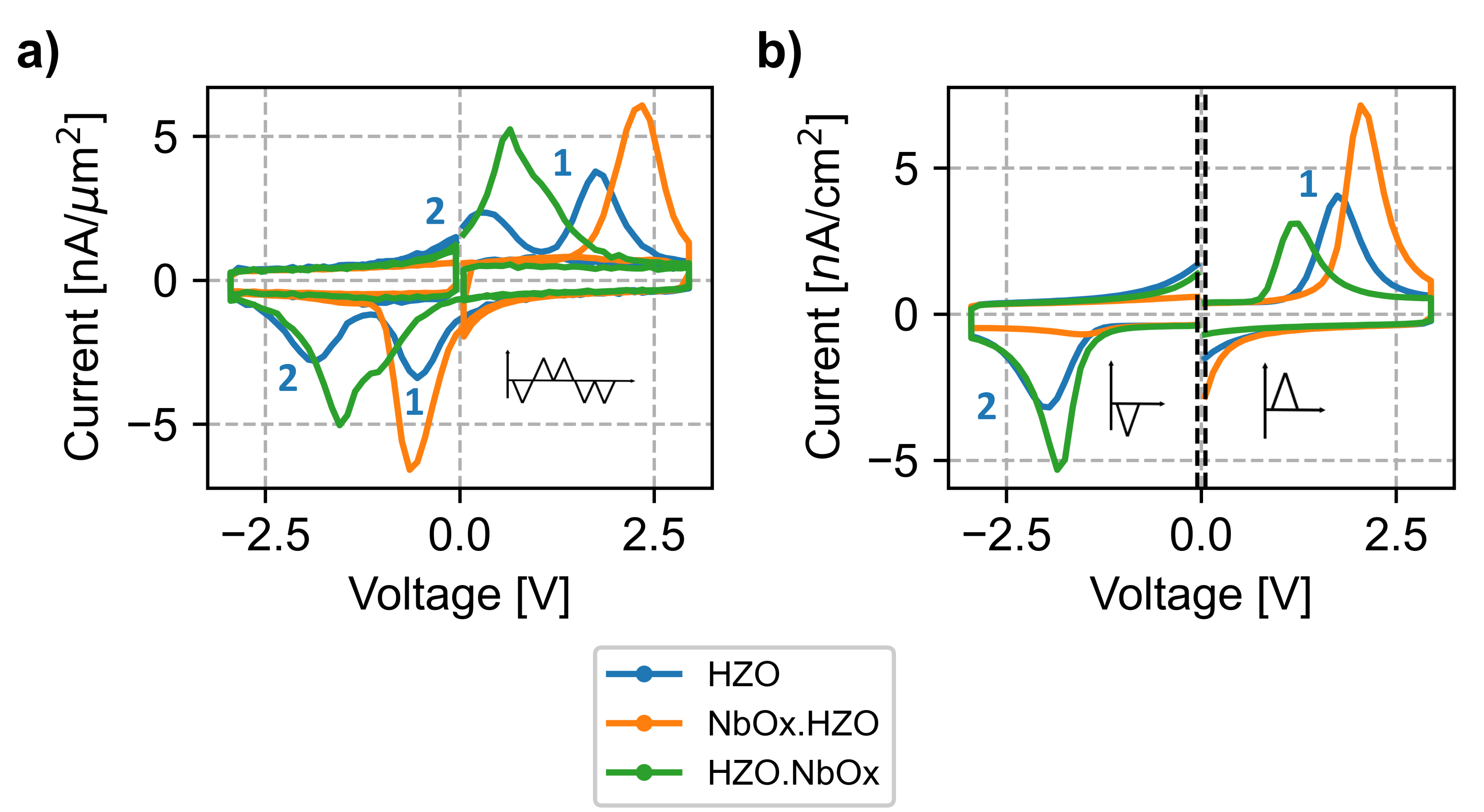}
\caption{PUND measurements for the pristine devices are shown in a) with the assignment of switching peaks to domain types for the HZO device. Before the PUND measurement a pre-pulse is used to bring the domains in the same state. The IV-curves measured with the application of either a positive or a negative pulse on two distinct pristine devices are shown in b). Blue shows the HZO sample, orange the NbO\textsubscript{x}.HZO sample and green the HZO.NbO\textsubscript{x }sample. }
\label{fig:pristine_MFM}
\end{figure}

The IV- and PV-curves measured with increasing amount of bipolar cycling of the MFM sample and the MFM samples with a NbO\textsubscript{x} layer inserted are shown in Fig. \ref{fig:IV_MFM}. \\
The device without NbO\textsubscript{x} shows strong peak splitting in the initial state, which is reduced with increasing amount of cycling. In the pristine HZO device, some of the domains are polarized in the P\textsubscript{up} state, marked with 1, and some are in the down state, marked with 2, as shown in Fig. \ref{fig:pristine_MFM} a).
Applying a positive pulse to one device and a negative pulse to an adjacent device causes the two types of domains to appear separately in the measured IV curves, as shown in Fig. \ref{fig:pristine_MFM} b). The switching peaks show an imprint in opposite directions, which could be a result of accumulated charges at opposing boundaries of the HZO, depending on the polarization state. \\
The introduction of NbO\textsubscript{x} near the top electrode in the HZO.NbO\textsubscript{x} device results in more domains of type 2 (average polarization pointing towards the bottom electrode) in the pristine state. Additionally, only domains of type 2 show an imprint to negative voltages, while domains of type 1 show no imprint as the peak position in Fig. \ref{fig:IV_MFM} c) indicates.\\
In the third case, with NbO\textsubscript{x} at the bottom interface in the NbO\textsubscript{x}.HZO device, almost all domains are polarized in the state 1 in the pristine state (average polarization pointing towards the top electrode) and have the same hysteresis shift to positive voltages caused by the internal electric bias field. As a consequence, no peak splitting is observed in this device.\\
The introduction of the NbO\textsubscript{x} layer significantly alters the mean average polarization direction in the pristine state and therefore the initial distribution of domains with an up- or downward projected polarization vector. 
The opposing average bias shift in the HZO.NbO\textsubscript{x} (shift to negative voltages) and NbO\textsubscript{x}.HZO (shift to positive voltages) samples shows that the bias shift depends on the position of the NbO\textsubscript{x} layer in the stack. The bias shift is most likely caused by accumulated charges in the NbO\textsubscript{x} layer or it's neighbouring interfacial regions, which could either emanate from charged trap sites \cite{zhang2024quantification} or the presence of charged oxygen vacancies in the HZO. The XPS measurements have shown an increased amount of vacancies in the HZO when it is in direct contact with the NbO\textsubscript{x}, but an influence of trap sites can't be excluded.\\
The influence of the NbO\textsubscript{x} layer on the distribution of domain types 1 and 2 is presumably more dominant in the NbO\textsubscript{x}.HZO sample than in the HZO.NbO\textsubscript{x} sample, because of either a higher chemical interaction between HZO and NbO\textsubscript{x} during the ALD process or due to the increased amount of oxygen in the NbO\textsubscript{x} layer of the NbO\textsubscript{x}.HZO sample as shown in Fig. \ref{fig:diffusion}.\\
Furthermore, the formation of alloys between HZO and NbO\textsubscript{x} can't be excluded, as it is already shown by P. Luo et al. \cite{luo2022phase} for ZrO\textsubscript{2}–Ta\textsubscript{2}O\textsubscript{5}–Nb\textsubscript{2}O\textsubscript{5} systems.\\
Bipolar cycling removes the accumulated vacancies from the interfaces and redistributes them in the HZO layer \cite{lee2023analysis,starschich2016evidence,pevsic2016physical}, which results in the observed reduction of the peak splitting and reduction of the bias shift with cycling\cite{lee2023analysis}.

\section{Electrical analysis of MFIM with NbO\textsubscript{x} incorporation}
\begin{figure}[H]
\includegraphics{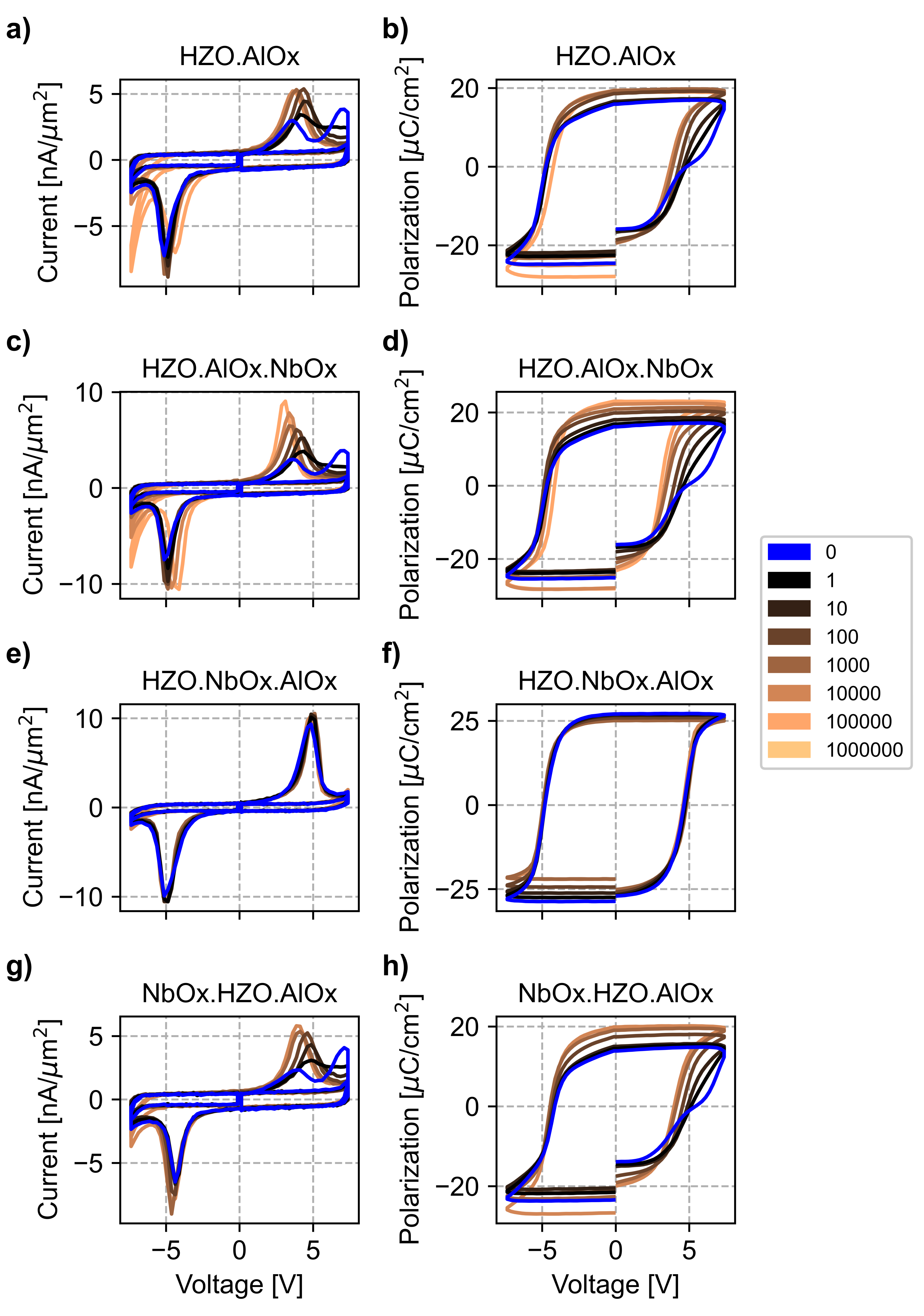}
\caption{IV- and PV-curves from PUND measurement with increasing cycling number for the investigated HZO.AlO\textsubscript{x} sample in a) and b), the HZO.AlO\textsubscript{x}.NbO\textsubscript{x} sample in c) and d), the HZO.NbO\textsubscript{x}.AlO\textsubscript{x} sample in e) and f), as well as the NbO\textsubscript{x}.HZO.AlO\textsubscript{x} sample in g) and h). For the IV-curves the current densities vs the applied voltages are plotted and for the PV loops the difference of P-U and N-D pulses are used to estimate the polarization.}
\label{fig:IV_MFIM}
\end{figure}

\begin{figure}[H]
\includegraphics{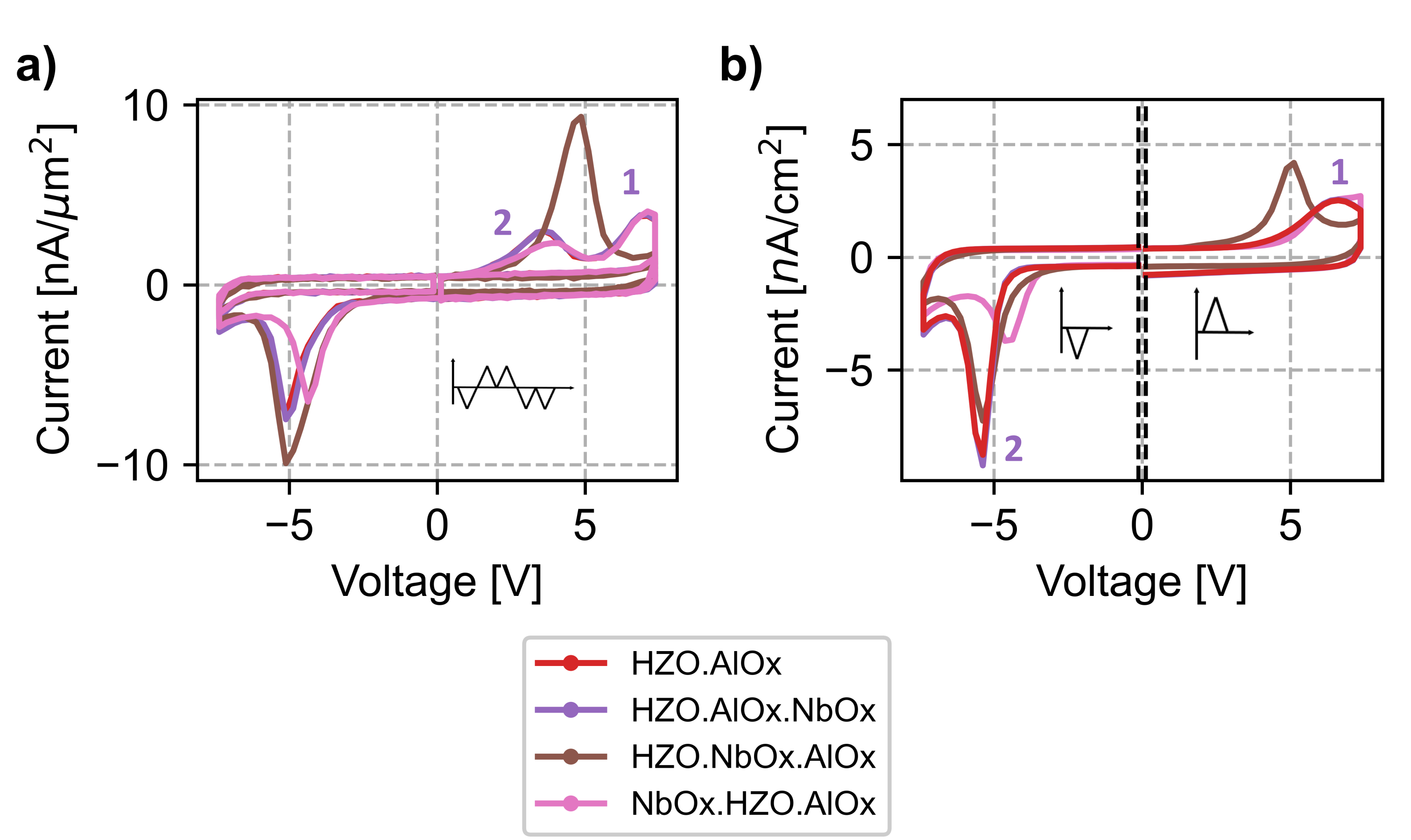}
\caption{PUND measurements for the pristine devices are shown in a) with the assignment of switching peaks to domain types for the HZO device. Before the PUND measurement a pre-pulse is used to bring the domains in the same state. The IV-curve for the application of a positive and negative pulse at two separate, pristine devices is shown in b). Red shows the HZO.AlO\textsubscript{x}, purple the HZO.AlO\textsubscript{x}.NbO\textsubscript{x}, brown the HZO.NbO\textsubscript{x}.AlO\textsubscript{x} and pink the NbO\textsubscript{x}.HZO.AlO\textsubscript{x}}
\label{fig:pristine_MFIM}
\end{figure}

The measured IV-curves and PV-curves of the MFIM sample and of the three MFIM samples with an inserted NbO\textsubscript{x} layer are shown in Fig. \ref{fig:IV_MFIM}.\\
All samples exhibit similar behavior except for the HZO.NbO\textsubscript{x}.AlO\textsubscript{x} one. Peak splitting is observed for positive voltages, but the second peak at a higher voltage (>5\,V) can not be fully measured due to the breakdown of the device. Only one peak is observed for negative voltages, which may arise from leakage mediated switching due to charge injection \cite{si2019ferroelectric,fontanini2022interplay} through the AlO\textsubscript{x}. The observed peak splitting consistently decreases with cycling, similar to what is observed in the MFM samples. The two switching peaks can be separated into domains of type 1 (P\textsubscript{up} domains in the pristine state with positive bias shift) and type 2 (P\textsubscript{down} domains with negative bias shift in the pristine state) as shown in Fig. \ref{fig:pristine_MFIM} a) and b).\\
The incorporation of NbO\textsubscript{x} below the HZO layer in the NbO\textsubscript{x}.HZO.AlO\textsubscript{x} device results in a shift of the IV curve to positive voltages, which is consistent with the observed behavior for the NbO\textsubscript{x}.HZO device. However, the device still shows peak splitting in the pristine state.\\
The HZO.AlO\textsubscript{x}.NbO\textsubscript{x}  sample shows no bias shift of the IV-curve in the pristine state (Fig. \ref{fig:pristine_MFIM} (a)) in comparison to the HZO.AlO\textsubscript{x} device, because there is no direct contact between the HZO and the NbO\textsubscript{x} layer and therefore no additional vacancies are generated in the HZO for this sample. Consequently, the behaviors of HZO.AlO\textsubscript{x} and HZO.AlO\textsubscript{x}.NbO\textsubscript{x} in the pristine state are very similar. After bipolar cycling, the devices show a deviation in their behaviour.\\
The device HZO.NbO\textsubscript{x}.AlO\textsubscript{x} shows almost no wakeup, no peak splitting in the pristine state and a very symmetric IV-curves as shown in Fig.\ref{fig:IV_MFIM}. However, the endurance of the device is reduced to 1e4 in comparison to 1e5 cycles for the other MFIM devices with AlO\textsubscript{x} layer. The lack of imprint and peak splitting for positive voltages indicates that the switching mechanism for positive voltages is driven by leakage as in the case for negative voltages.

In the study by Si et. al. \cite{si2019ferroelectric} a charge trapping induced switching model is used to describe the switching behavior of HZO/AlO\textsubscript{x} stacks, which could explain the reduced effect of NbO\textsubscript{x} in the NbO\textsubscript{x}.HZO.AlO\textsubscript{x} device compared to the one in the NbO\textsubscript{x}.HZO device. The observed device behavior is consequently not entirely dominated by the generated vacancies of the NbO\textsubscript{x} layer, but by charge injection processes over the AlO\textsubscript{x} layer. \\
Assuming a series capacitance of HZO and AlO\textsubscript{x} capacitors, with a 4\,nm AlO\textsubscript{x} and 10\,nm HZO thicknesses and a dielectric constanst of 30 for the HZO and 7 to 9 for AlO\textsubscript{x}, then the switching voltage of the device should be between 2.7 and 3.4\,V, assuming a necessary voltage drop of 1.2\,V over the HZO layer for switching the domains.\\
Nevertheless, the observed switching peaks for the HZO.AlO\textsubscript{x} device in the cycled state are around $\pm$5\,V and is higher than the calculated voltage with the series capacitance model. It can be assumed that the device behavior is strongly influenced by charge traps at the HZO-/AlO\textsubscript{x} interface and that the charge injection mechanism plays a crucial role in the device properties such as switching fields. The asymmetric IV curve of the HZO.AlO\textsubscript{x} device could be due to different dominant charge injection mechanism for positive and negative voltages. For the HZO.NbO\textsubscript{x}.AlO\textsubscript{x} device, the charge injection mechanism is modified by the NbO\textsubscript{x} layer, resulting in a symmetric IV curve. 

\section{Conclusion}

The experiments show that the insertion of a NbO\textsubscript{x} layer in MFM devices based on hafnium zirconium oxide modifies the pristine device properties such as imprint and peak-splitting when the NbO\textsubscript{x} is in direct contact with the HZO layer.
The NbO\textsubscript{x} scavenges oxygen from the neighboring HZO layer and creates double positively charged oxygen vacancies in the HZO, which are likely the cause of the observed imprint when NbO\textsubscript{x} is introduced into the MFM stack. \\
A stronger influence of the NbO\textsubscript{x} layer is observed in NbO\textsubscript{x}.HZO devices compared to HZO.NbO\textsubscript{x} ones, which is most likely related to the increased amount of oxygen in the NbO\textsubscript{x} layer as shown in EELS measurements and thus the exposure of the NbO\textsubscript{x} layer in the NbO\textsubscript{x}.HZO sample to the ALD process. \\
In addition, inserting the NbO\textsubscript{x} layer into the stack causes a preferred polarisation state whose direction depends on the position of the NbO\textsubscript{x} in the stack. The preferred polarisation state is most likely related to the electric bias field caused by the charged oxygen vacancies at the HZO-/NbO\textsubscript{x} interface. 
The influence of the introduced NbO\textsubscript{x} layer causes similar bias shifts in the pristine device state as previously reported for the incorporation of thin tantalum oxide layer\cite{jeong2022oxygen,shin2023method} in HZO-based MFM capacitors. Unlike MFM stacks with a tantalum oxide layer, where the bias shift is caused by the creation of fixed charges in the tantalum oxide, the bias shift in our devices reduces with cycling, which is most likely caused by a redistribution of vacancies in the HZO layer\cite{lee2023analysis,pevsic2016physical}.\\
The introduction of a NbO\textsubscript{x} layer in MFIM stacks based on a stack of HZO and AlO\textsubscript{x} shows the expected bias shift in the NbO\textsubscript{x}.HZO.AlO\textsubscript{x} sample, but the influence is not as strong as in the NbO\textsubscript{x}.HZO sample.
A possible explanation is the influence of charge trapping induced switching at the HZO-AlO\textsubscript{x} interface. The switching process is then at least partially defined by the charge trapping process, which can explain the asymmetric device behavior and the deviation of the measured switching voltages and the switching voltages extracted from a series capacitance model. 
The incorporation of NbO\textsubscript{x} in between HZO and AlO\textsubscript{x} in the HZO.NbO\textsubscript{x}.AlO\textsubscript{x} device leads to a symmetric IV curve. A likely cause is the change of the AlO\textsubscript{x} layer by the adjacent NbO\textsubscript{x} layer. Further experiments are necessary to identify the charge injection mechanism and its modification by the incorporation of NbO\textsubscript{x} in the HZO.NbO\textsubscript{x}.AlO\textsubscript{x} device. 

\newpage
\begin{acknowledgement}
M.E. acknowledges funding from the German Research Foundation as part of the FeDiBiS project with project ID 449644906.\\
C.D., S.S.  acknowledge funding from the European Union’s Horizon research and innovation programme under grant agreement 101135398 (FIXIT). \\
L.C. acknowledges funding from the European Union's European Innovation Council under Grant Agreement No. 101070908 (CROSSBRAIN).\\
S.L. acknowledges funding from European Union’s Horizon research and innovation programme (Grant agreement No. 101135656) Ferro4EdgeAI\\
C.D. and I.H. acknowledge funding by the European Union of the project number 101098216 (ERC Advanced Grant, LUCIOLE,). Views and opinions expressed are, however, those of the authors only and do not necessarily reflect those of the European Union or the European Research Council Executive Agency. Neither the European Union nor the granting authority can be held responsible for them.\\
T.M. acknowledges funding from the Saxonian State budget approved by the delegates of the Saxon State Parliament.

\end{acknowledgement}

\newpage
\begin{suppinfo}
The supplementary information contains additional information about electric, EELS and XPS measurements.\\
A description of how to generate the cross section shown in Fig. \ref{fig:TEM} is given.\\
Furthermore, XPS measurements on the Hf 3d core level of the device HZO.NbO\textsubscript{x}.AlO\textsubscript{x} as well as XPS measurements on the Al 2p core level for the devices HZO.AlO\textsubscript{x} and HZO.NbO\textsubscript{x}.AlO\textsubscript{x} are shown.\\
The electrical measurement setup, the electrical parameters of the samples used in XPS, and the P\textsubscript{r} development of the investigated MFM and MFIM samples with increasing bipolar cycling are shown.\\
In addition, the IV development with cycling for different voltages of the HZO.NbO\textsubscript{x}.AlO\textsubscript{x} device is shown.

\end{suppinfo}
\newpage

\bibliography{main}

\end{document}

\newpage
\subsection{TEM image}

\begin{figure}[H]
\includegraphics[scale=0.8]{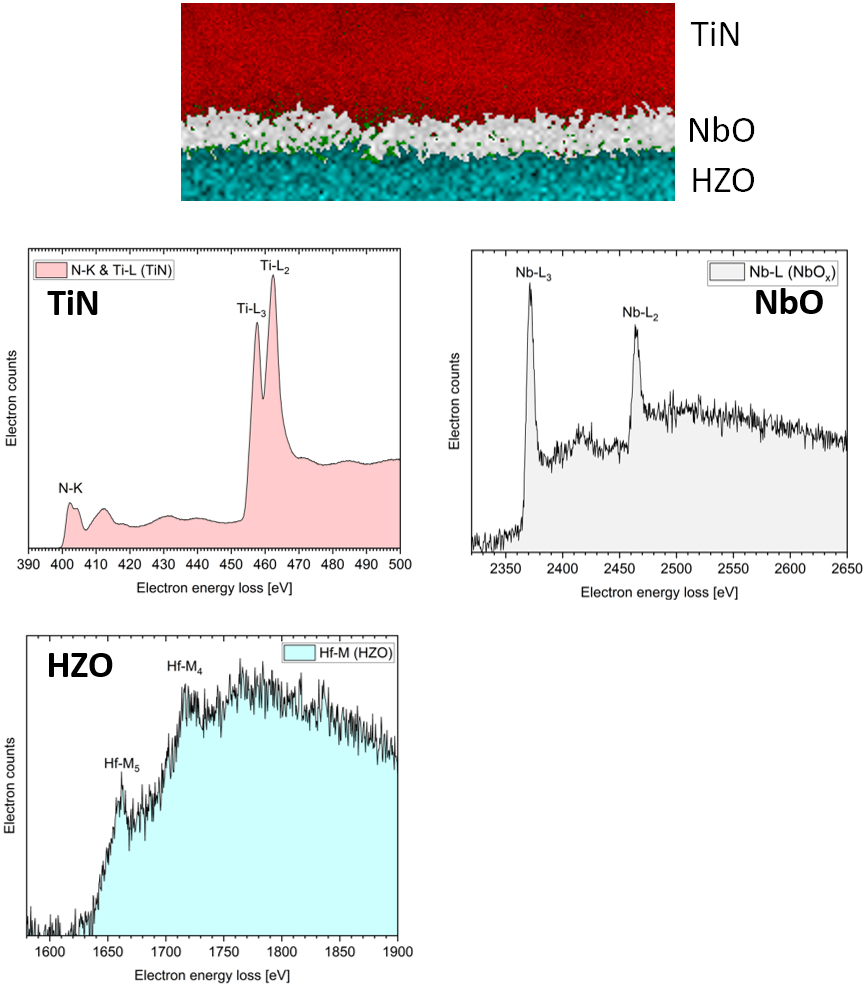}
\caption{Overlay of EELS measurements for the different layers. Each color-dot in the cross section represents the occurrence of the signal for the specific material. For red dots the Ti signal, for grey the Nb signal and for blue the Hf signal is used.}
\label{fig:tem_extraction}
\end{figure}

The image extraction for the TEM cross section of the materials is shown in Fig. \ref{fig:tem_extraction}. 

\newpage
\subsection{Extraction of XPS data}

The composition of NbO\textsubscript{x} is calculated by analyzing the Nb 3d core level. The measurement results would indicate that 100\% of the Nb is in the Nb\textsuperscript{5+} state, which means that all of the NbO\textsubscript{x} is in the stoichiometry Nb\textsubscript{2}O\textsubscript{5}. The deviation between fit and measured data, which can be caused by various physical mechanism as outlined in the manuscript, indicates that the extracted value might be incorrect and that the extraction of the Nb\textsubscript{2}O\textsubscript{5} stoichiometry is beyond sensitivity. 

\begin{figure}[H]
\includegraphics[scale=0.6]{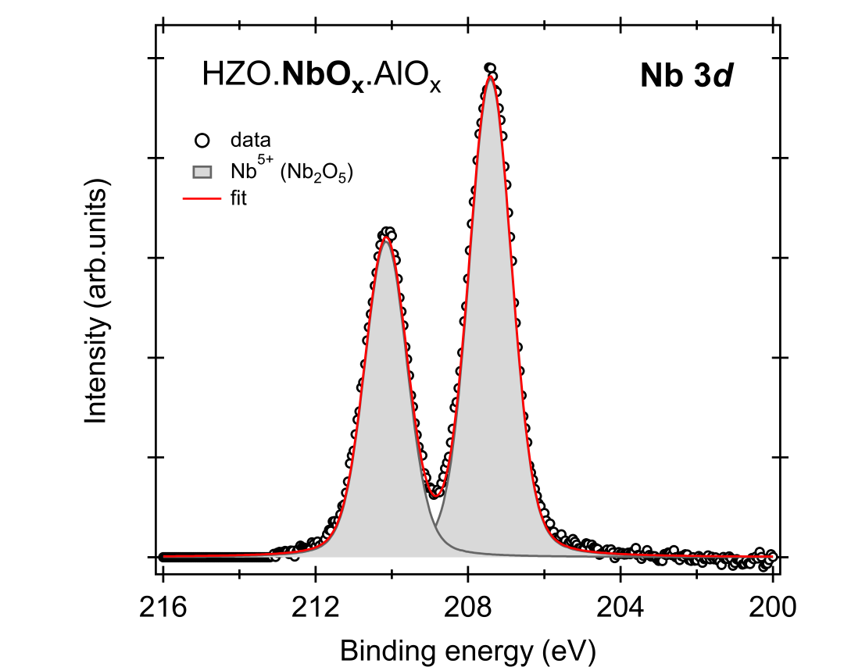}
\caption{XPS measurement on the Hf 3d core level for the HZO.NbO\textsubscript{x}.AlO\textsubscript{x} device. The red curve is the fit of the experimental data.}
\label{fig:nbo_composition}
\end{figure}

\newpage
The composition of the aluminum oxide layer is calculated by analyzing the Al 2p core level. For the simulation it is assumed that only phases in the Al\textsubscript{2}O\textsubscript{3} composition and pure metallic composition are present. 

\begin{figure}[H]
\includegraphics[scale=0.7]{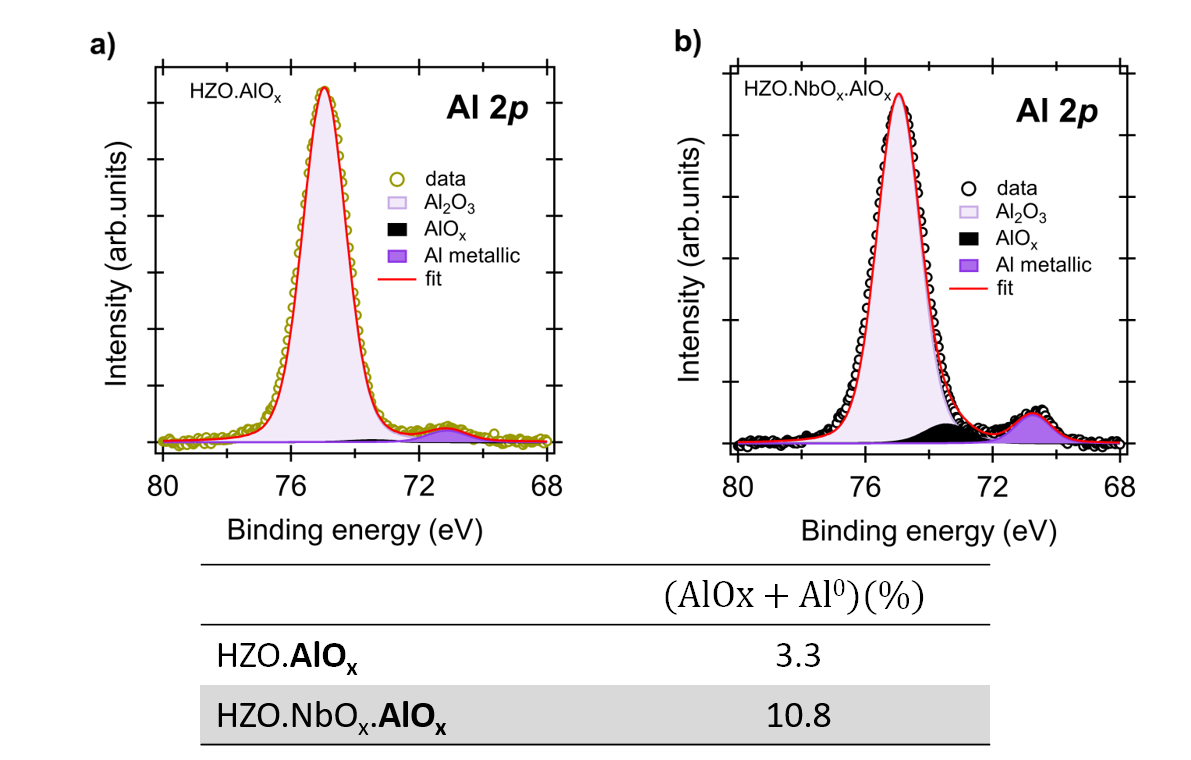}
\caption{XPS measurement on the Al 2p core level for the HZO.AlO\textsubscript{x} device in a) and for the HZO.NbO\textsubscript{x}.AlO\textsubscript{x} device in b). The red curve is the fit of the experimental data by assuming the presence of Al\textsubscript{2}O\textsubscript{3} and metallic compositions in the aluminium oxide layer. The black curve is the difference between the fit and the measured data and represents the amount of Al not fitting in one of the two chemical compositions and thus representing substoichiometric AlO\textsubscript{x}. In the table below is the extracted peak intensity of metallic Al and AlO\textsubscript{x} in relation the measured peak intensity of the Al 2p core level shown.}
\label{fig:alox_composition}
\end{figure}

\newpage
\subsection{Electrical measurements on XPS samples}

\begin{figure}[H]
\includegraphics{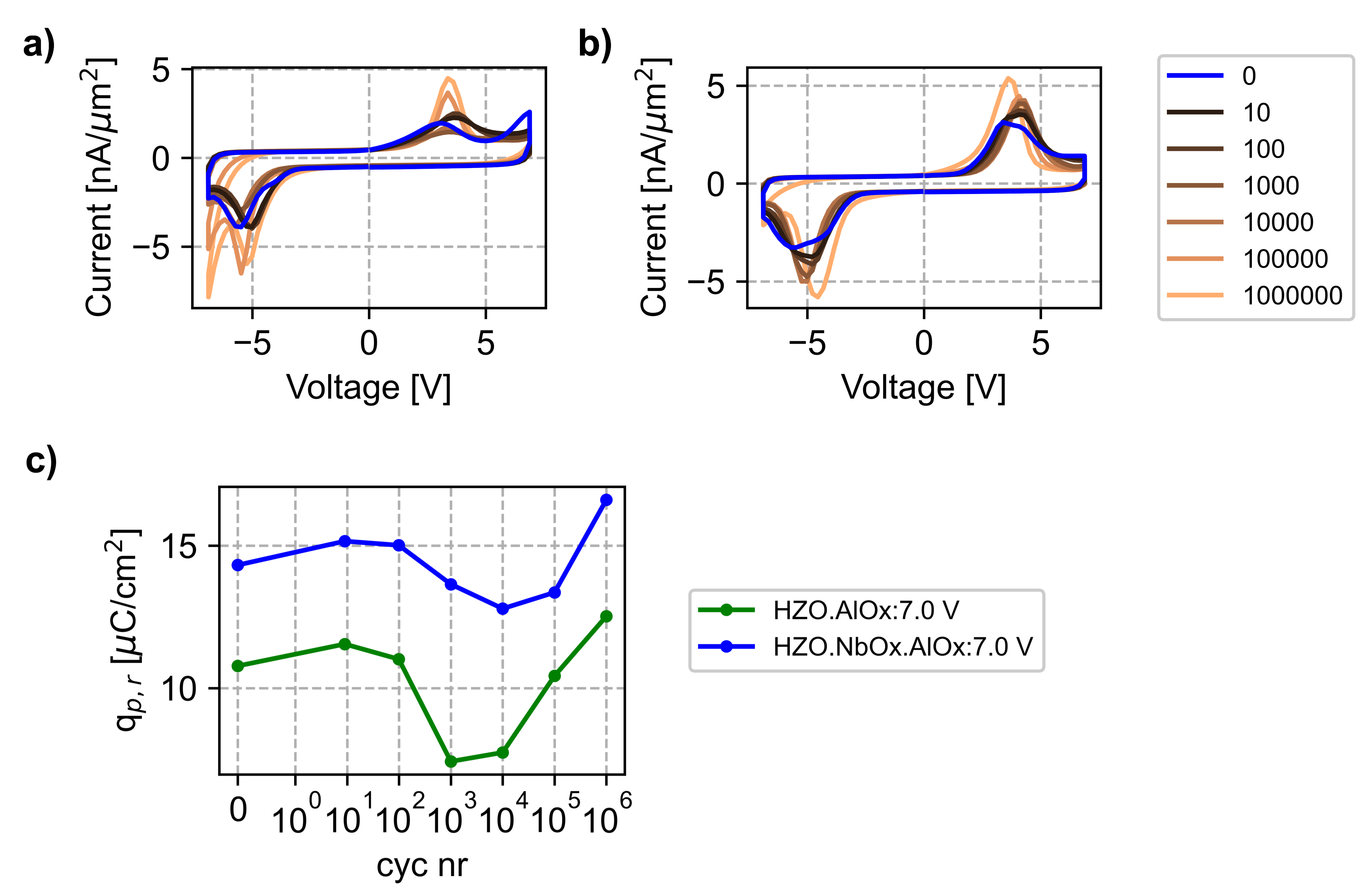}
\caption{IV-curve development with cycling at 7\,V for the HZO.AlO\textsubscript{x} device in a) and HZO.NbO\textsubscript{x}.AlO\textsubscript{x} device in b). The charge of the P-U pulse with increasing number of cycles for 7\,V is shown in c) for the HZO.AlO\textsubscript{x} device in green and the HZO.NbO\textsubscript{x}.AlO\textsubscript{x} device in blue.}
\label{fig:electric measurement xps}
\end{figure}

Electrical measurements performed on the devices for XPS measurements are shown in Fig. \ref{fig:electric measurement xps}. The Pr is higher and the peak-splitting is reduced for the HZO.NbO\textsubscript{x}.AlO\textsubscript{x} sample in comparison to the HZO.AlO\textsubscript{x} sample.

\subsection{Measurement setup}

\begin{figure}[!htb]
\includegraphics[width=6.5in]{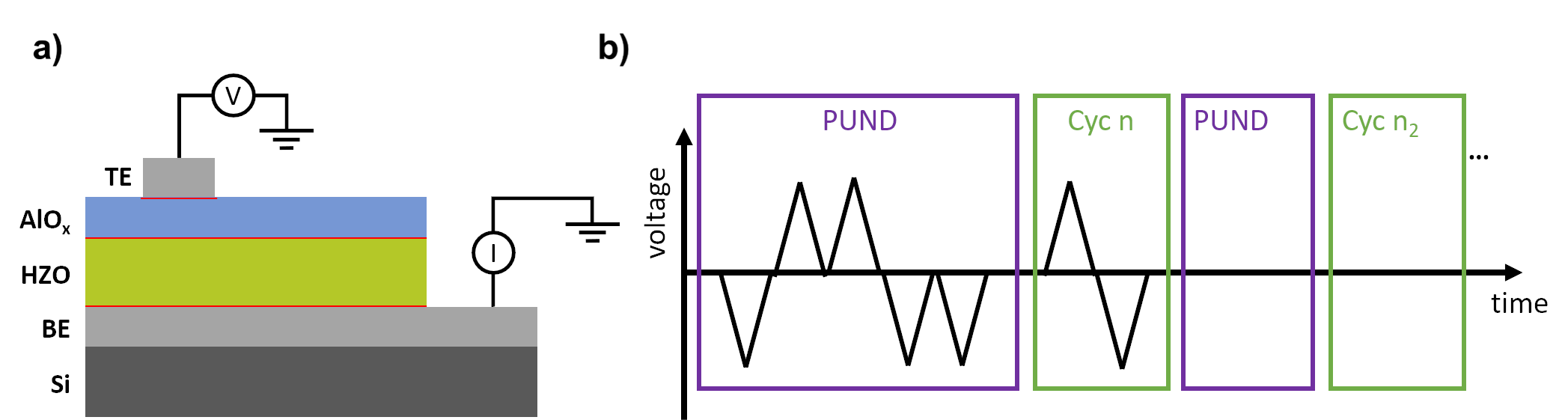}
\caption{Measurement setup illustrated with a MFIM device in a) and used pulse sequence for IV-curve extraction in b). The frequency of the PUND measurement is 1\,kHz and for the cycling 10\,kHz. The amplitude for the measurement and cycling pulses is set to 3\,V for devices without AlO\textsubscript{x} layer and 7.5\,V for devices with AlO\textsubscript{x} layer.}
\label{fig:setup_sequence}
\end{figure}

The measurement setup used for all devices is shown in Fig. \ref{fig:setup_sequence} a) with the cross section of a MFIM sample with HZO as ferroelectric and AlO\textsubscript{x} as insulating layer. The used pulse sequence for electric characterization is depicted in Fig. \ref{fig:setup_sequence} b).

\newpage
\subsection{Electrical Measurement data}

\begin{figure}[H]
\includegraphics{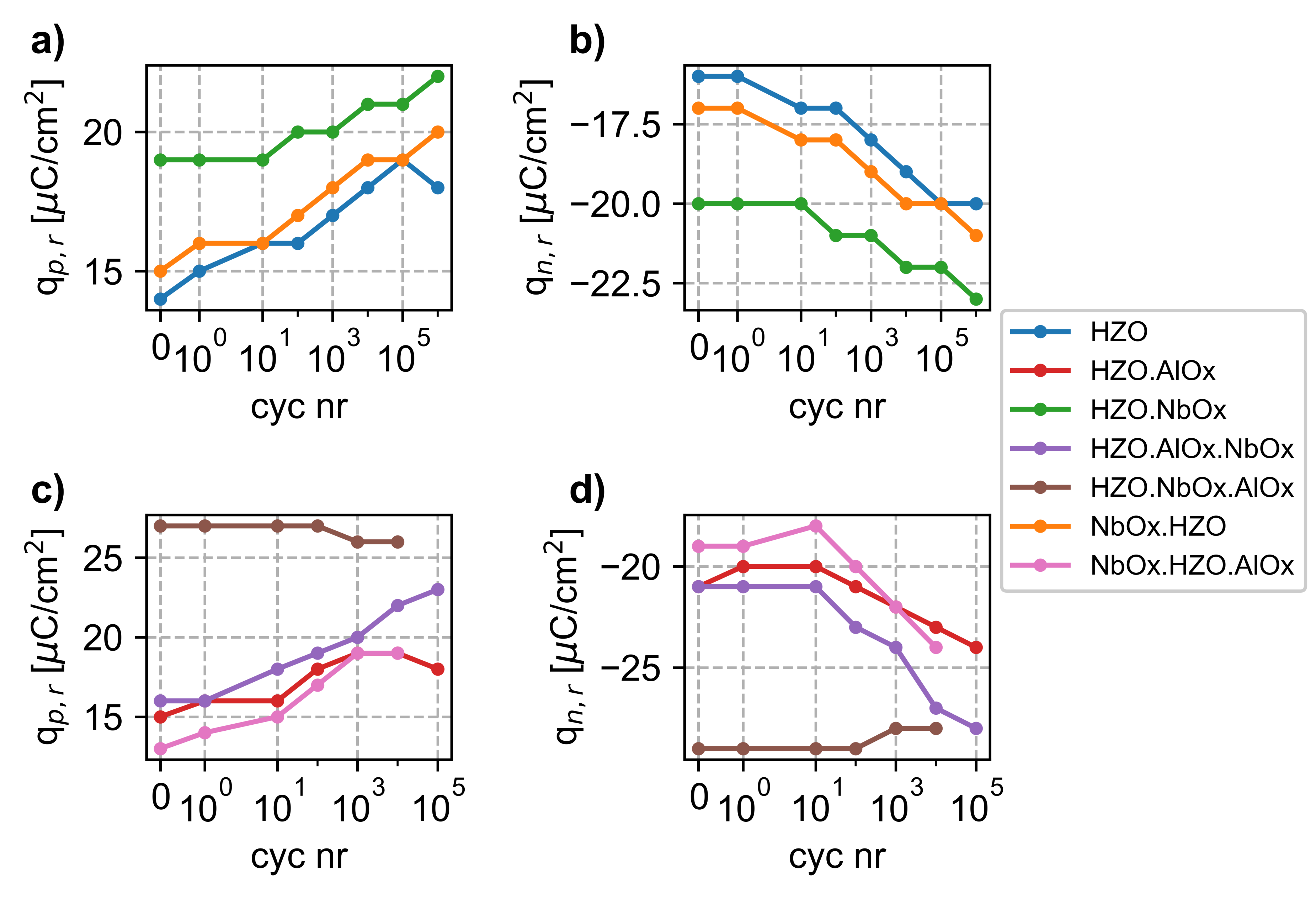}
\caption{Measured remanent charge from the difference between P- and U- pulse vs number of cycling in a) and c) and from the difference between N- and D- pulse in b) and d). 3\,V is used for the measurements of the samples without AlO\textsubscript{x} and 7.5\,V is used for the samples with AlO\textsubscript{x}. Each color represents one sample type as shown in the legend.}
\label{fig:Pr}
\end{figure}

Measurement of the polarizations with increasing amount of bipolar cycling for the samples used in TEM images and for electric characterization in Fig \ref{fig:Pr}.

\begin{figure}[H]
\includegraphics{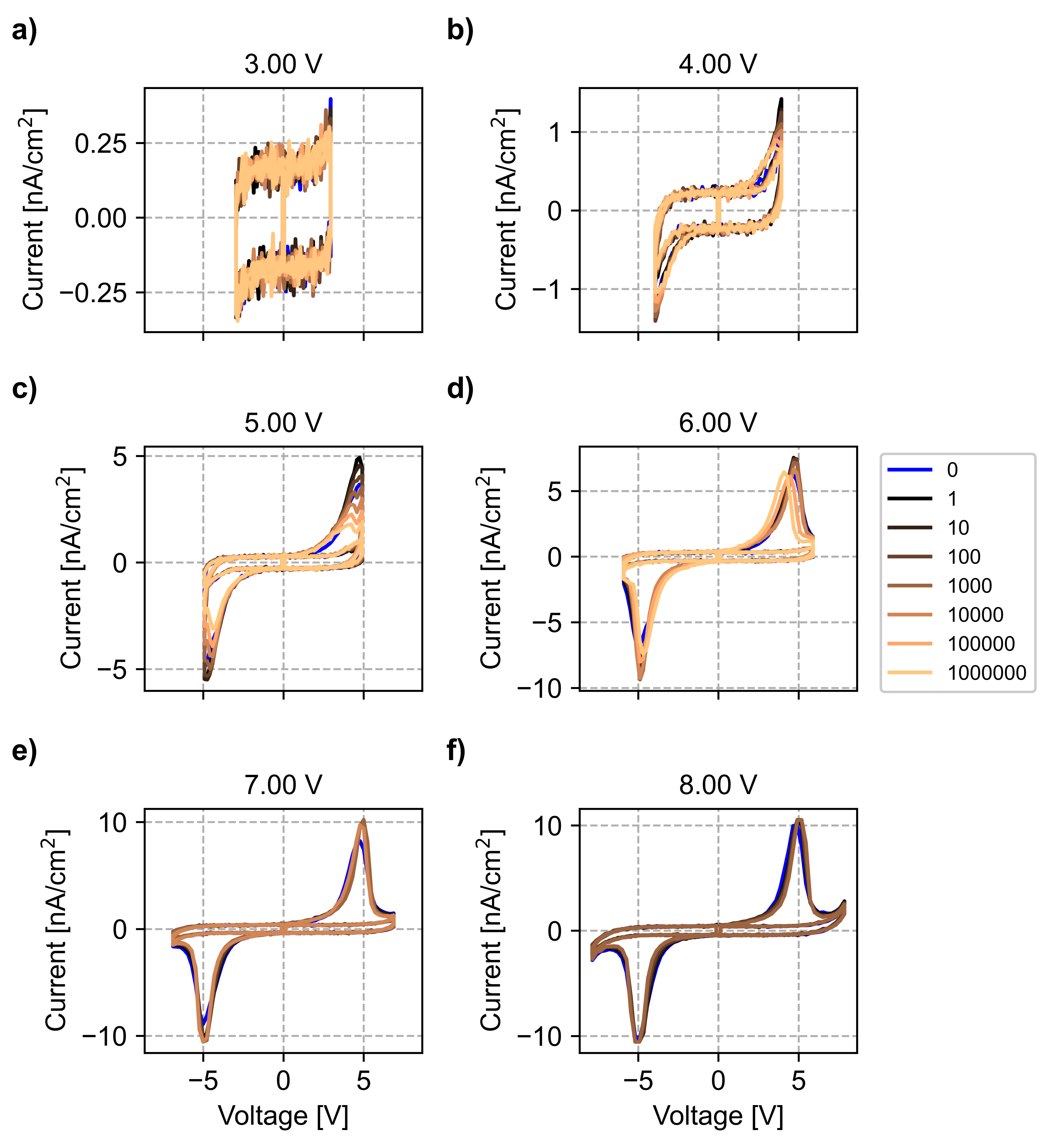}
\caption{Measured IV curves of the HZO.NbO\textsubscript{x}.AlO\textsubscript{x} sample with increasing voltages from a) to f) as indicated by the subplot title. The voltage was used for cycling pulses and measurement pulses. The colors represent the number of cycles as shown in the legend.}
\label{fig:IV_HZONbOAlO_appendix}
\end{figure}

\newpage
The IV curves in Fig. \ref{fig:IV_HZONbOAlO_appendix} show that a stable IV behavior in the HZO.NbO\textsubscript{x}.AlO\textsubscript{x} sample over cycling is only achieved for cycling and measurement voltages that encompass the entire switching peak. 

